\documentclass[amssymb,aps,prb,twocolumn,showpacs,superscriptaddress]{revtex4}
\usepackage{graphicx}
\usepackage{dcolumn}
\usepackage{bm}

\begin{document}

\title{Magnetization reversal and anomalous coercive field temperature dependence
in MnAs epilayers grown on GaAs(100) and GaAs(111)B}
\author{L. B. Steren}
\affiliation {Centro At\'omico Bariloche, CNEA-UNC, (8400) S.C. de
Bariloche, Argentina.}
\author{J. Milano}
\affiliation {Centro At\'omico Bariloche, CNEA-UNC, (8400) S.C. de
Bariloche, Argentina.}
\author{V. Garcia}
\affiliation {Institut des NanoSciences de Paris, INSP, Universit\'e
Pierre et Marie Curie-Paris 6, Universit\'e Denis Diderot-Paris 7,
CNRS UMR 7588, Campus Boucicaut, 140 rue de Lourmel, 75015 Paris,
France.}
\author{M. Marangolo}
\affiliation {Institut des NanoSciences de Paris, INSP, Universit\'e
Pierre et Marie Curie-Paris 6, Universit\'e Denis Diderot-Paris 7,
CNRS UMR 7588, Campus Boucicaut, 140 rue de Lourmel, 75015 Paris,
France.}
\author{M. Eddrief}
\affiliation {Institut des NanoSciences de Paris, INSP, Universit\'e
Pierre et Marie Curie-Paris 6, Universit\'e Denis Diderot-Paris 7,
CNRS UMR 7588, Campus Boucicaut, 140 rue de Lourmel, 75015 Paris,
France.}
\author{V. H. Etgens}
\affiliation {Institut des NanoSciences de Paris, INSP, Universit\'e
Pierre et Marie Curie-Paris 6, Universit\'e Denis Diderot-Paris 7,
CNRS UMR 7588, Campus Boucicaut, 140 rue de Lourmel, 75015 Paris,
France.}

\begin{abstract}
The magnetic properties of MnAs epilayers have been investigated for
two different substrate orientations: GaAs(100) and GaAs(111). We
have analyzed the magnetization reversal under magnetic field at low
temperatures, determining the anisotropy of the films. The results,
based on the shape of the  magnetization loops, suggest a domain
movement mechanism for both types of samples. The temperature
dependence of the coercivity of the films has been also examined,
displaying a generic anomalous reentrant behavior at T$>$200 K. This
feature is independent of the substrate orientation and films
thickness and may be associated to the appearance of new pinning
centers due to the nucleation of the $\beta$-phase at high
temperatures.
\end{abstract}

\date{\today}
\maketitle

\section{INTRODUCTION}
In the last few years, a lot of effort has been invested to
understand the physical properties of MnAs-based thin films and
multilayers. Bulk manganese arsenide is ferromagnetic at room
temperature with a NiAs hexagonal structure ($\alpha$-phase).
\cite{Wilson,willis,goodenough} At $T_{C}=$313 K however, MnAs
displays a first-order phase transition that drives the system to a
paramagnetic state. At this temperature, a structural transition
occurs from the low-temperature $\alpha$-phase to an
orthorhombically-distorted (MnP) $\beta$-phase. At $T_{St}=$398 K, a
second structural transition takes place and the system regains a
NiAs hexagonal structure. MnAs grows epitaxially on various standard
semiconductor substrates such as GaAs(100),
\cite{Tanaka,Trampert,Chun} GaAs(111)B,
\cite{Tanaka2,Jenichen,Mattoso} GaAs(110), \cite{Kolovos} InP(100),
\cite{Yokoyama} Si(100) \cite{Akeura} and Si(111). \cite{Nazmul} The
excellent quality of MnAs/GaAs heterostructures makes this compound
particularly interesting for hybrid metal/semiconductor spintronics.
\cite{TanakaIII,Ramsteiner,ChunII,Garcia}

\begin{figure}[htb]
\includegraphics[width=\linewidth]{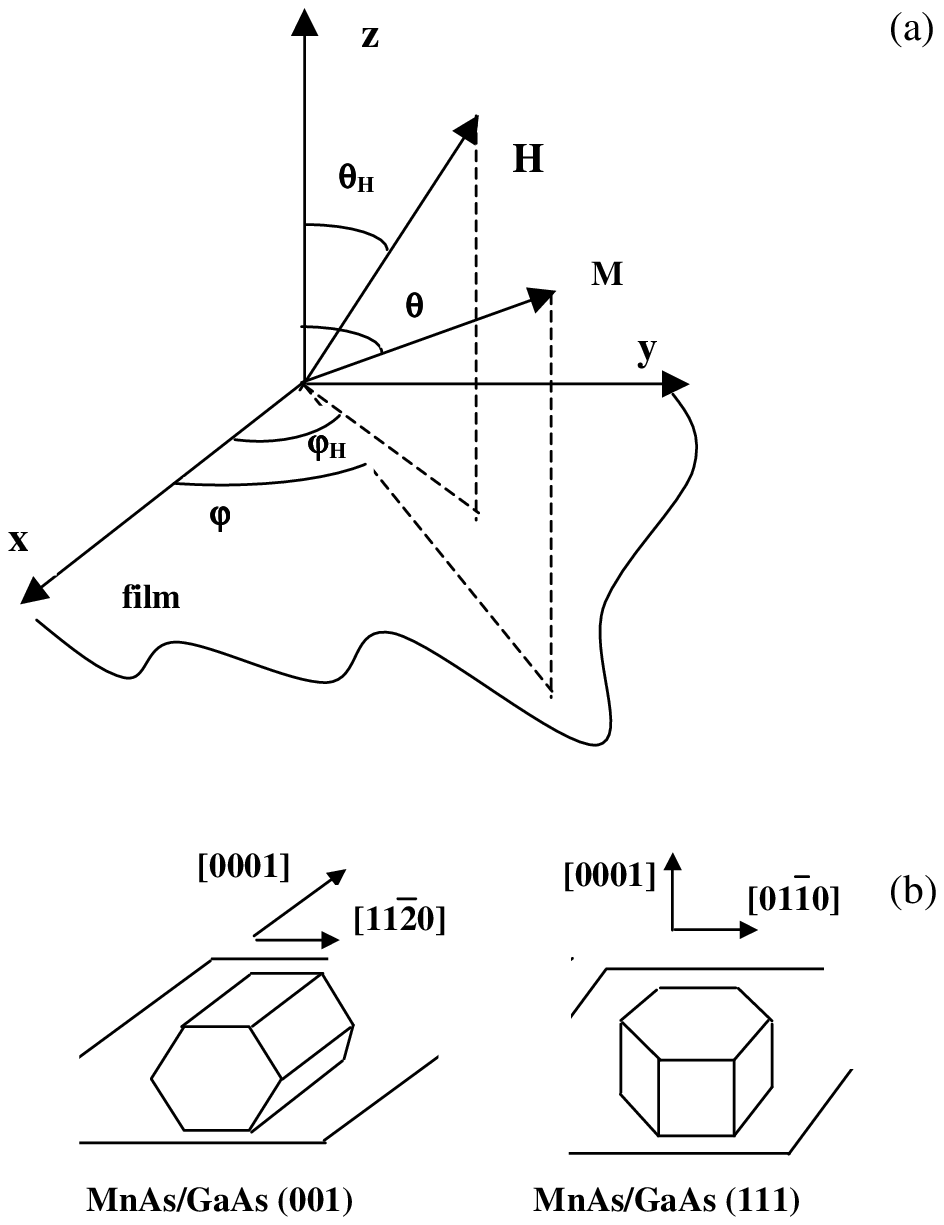}
\caption{(a) Coordinate system used in the paper; (b) Schematic view
of the epitaxial relationship of MnAs on GaAs(100) and GaAs(111),
respectively.}\label{Figure1}
\end{figure}

The epitaxy introduces strain and defects in MnAs thin films that
significatively modify their structural and magnetic properties with
repect to the bulk. When MnAs is grown on GaAs(100) or GaAs(111)B,
epitaxial strain induces an $\alpha/\beta$-phases coexistence in a
wide range of temperature. \cite{Das,Mattoso} In addition, the
$T_{C}$ of the thin films can be enhanced when they are grown on
GaAs(111)B substrates. \cite{Mattoso} The orientation of the GaAs
substrate determines the growth behavior and epilayer orientation.
When grown on GaAs(111)B substrate, MnAs lies with the hexagonal c
axis aligned with the growth direction [Fig. \ref{Figure1}(b)];
\cite{Tanaka2} while for the growth on GaAs(100) substrate, two
orientations of the epilayer have been described. \cite{Tanaka} By
correctly adjusting the growth conditions, it becomes possible to
obtain one single phase \cite{Schippan} with the MnAs $c$-axis
parallel to the GaAs [110] that lies inside the surface plane [Fig.
\ref{Figure1}(b)]. Due to the important misfit between the thin film
and the substrate (7$\%$), the epilayer relaxes its strain from the
very beginning of the growth. Above a certain thickness, the
epilayer is expected to be fully relaxed at the growth temperature.
However, during the after-growth cooling cycle, a residual strain is
introduced due to the large dilatation coefficients difference
between MnAs and GaAs. \cite{Iikawa} In fact, as the $\alpha-\beta$
phase transition proceeds mainly by a 1$\%$ lattice parameter
contraction in the hexagonal plane, strain will accommodate
differently for MnAs/GaAs(100) or MnAs/GaAs(111)B.

Another particularity of MnAs epilayers grown on GaAs(100)
substrates is that, within the phase coexistence region, scanning
tunneling microscopy (STM) images show the presence of stripes along
the [0001] direction with two different heights, arranged alternated
within the $[11\bar{2}0]$ direction. \cite{stripes} Magnetic force
magnetometry (MFM) studies indicate that the higher stripes are
ferromagnetic ($\alpha$-phase) while the shallower are paramagnetic
($\beta$-phase). \cite{Daweritz} X-ray diffraction (XRD) patterns
also agree with the $\alpha$-$\beta$ phase coexistence in the 280-
314 K temperature range. No stripes are however observed for MnAs
epilayers grown on GaAs(111)B. Instead, the surface morphology shows
triangular and sticks features depending on the growth parameters
like substrate temperature, As overpressure, etc. \cite{Ouerghi}
Even if no stripes are visible, the phase coexistence has been
unambiguously observed between 296 and 328 K by XRD for this
orientation. \cite{Mattoso}

In the present work, we perform a detailed study of the magnetism of
MnAs epilayers, grown on both GaAs(100) and GaAs(111)B substrate
orientations. The anisotropy of the films is analyzed in terms of
magneto-crystalline and substrate-induced anisotropy terms. The
temperature dependence of the coercivity of the films is measured
and explained in terms of thermally activated domain-wall movements
and the appearance of new pinning centers.

\section{EXPERIMENTAL DETAILS AND CHARACTERIZATION OF THE SAMPLES}

MnAs epilayers are grown by molecular beam epitaxy (MBE) with a
substrate temperature of about $240^{\circ}${C} on previously
prepared GaAs buffer layers.  After removal of the native oxide, a
very thin buffer layer is grown on GaAs(111)B substrates
\cite{GarciaII} while a standard 100 nm thick buffer layer is grown
on GaAs(100). Each MnAs growth begins under As rich conditions on
the GaAs buffer layer. Different epilayer thicknesses, $t_x$,
ranging from 30 nm to 200 nm, are grown on both (100) and (111)
substrate orientations. The films are named M100-$t$ [nm] and
M111-$t$ [nm], respectively. Finally, the samples are capped
\emph{in situ} with an amorphous ZnSe layer.

Magnetization vs. temperature and magnetic field curves are measured
in a vibrating sample magnetometer (VSM) and a superconducting
quantum interference device magnetometer (SQUID). The temperature is
varied from 5 K to 300 K and the magnetic fields swept up to 5 T.
The magnetization loops are measured with the magnetic field
rotating in the plane of the film (IP) $\theta_H= \pi/2$ and in the
out-of-plane geometry (OOP), keeping the azimuthal angle $\varphi_H$
fixed [Fig. \ref{Figure1}(a)].

\section{RESULTS AND DISCUSSION}
\begin{figure}[htb]
\includegraphics[width=\linewidth]{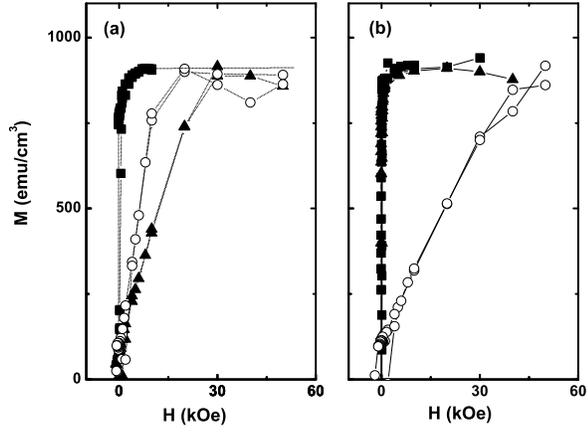}
\caption{Magnetization loops of (a) M100-66 and (b) M111-66 films
measured with ($\blacktriangle$) the magnetic field applied in the
plane of the films and perpendicular to the $x$-axis,
($\blacksquare$) in-plane and parallel to the $x$-axis and
($\bigcirc$) out-of- the plane of the films. The measurements were
performed at 5 K in the SQUID magnetometer.}\label{Figure2}
\end{figure}

The ferromagnetic phase with hexagonal structure displays a strong
anisotropy with a hard-axis parallel to the $c$-axis, that favors
the orientation of the magnetization in a plane perpendicular to
this direction. \cite{Deblois} As mentioned previously, the
hexagonal $c$-axis is inside the plane of the films for M100 samples
and perpendicular to it for M111 samples [Fig. \ref{Figure1}(b)],
which strongly affects the magnetism of the samples. The
magnetization loops reflect the structural stacking of MnAs in the
films. Fig. \ref{Figure2} shows the typical magnetization loops for
M100 and M111 samples. These have been measured in three different
configurations of the applied field: in-the-plane of the films and
parallel to the $x$-axis, IP-PA ($\theta_H = \pi/2$, $\varphi_H =
0$), in-plane and perpendicular to the $x$-axis, IP-PE ($\theta_H =
\pi/2$, $\varphi_H = \pi/2$) and out-of-the-plane of the films with
$\theta_H=0$ . The $x$-axis is parallel to the [11$\bar{2}$0] axis
for M100 samples, and is not clearly determined for M111 samples.

Fig. \ref{Figure2}(a) evidences the existence of a strong uniaxial
anisotropy in the plane of M100 films, with the easy-axis oriented
along the MnAs $[11\bar{2}0]$ direction. The saturation field of the
in-plane curve, measured with the magnetic field perpendicular to
the easy-axis direction, is much larger than the OOP one, indicating
that the anisotropy term overcomes the demagnetization effect in
these films. In M111 films [Fig. \ref{Figure2}(b)] and due to the
structure stacking onto the substrate, the hard axis strengthens the
demagnetization effect favoring an almost isotropic in-plane
magnetization.

The free energy density proposed is:
\begin{equation}
F_{100} = - \vec{M}.\vec{H} + K_1 \sin^2\theta.\sin^2\varphi
 + K_n\cos^2\theta+ 2\pi M_S\cos^2\theta
\end{equation}
\begin{equation}
F_{111} = - \vec{M}.\vec{H} + K_1 \cos^2\theta + 2\pi M_S
\cos^2\theta +  {\cal O}^2 (K_2 (\varphi))
\end{equation}
for M100 and M111 samples, respectively.

The first term in each equation accounts for the Zeeman energy and
the second term includes a uniaxial anisotropy oriented in the plane
of the film or out of the plane, depending on the samples substrate
orientation. The last term represents the magnetostatic energy.  To
describe the experimental results we have included an extra
anisotropy term in the M100 energy expression, $K_n$, that favors an
in-plane magnetization. We associate this contribution to
substrate-induced strains onto the film structure. In Ref. 18 A.K.
Das and coworkers already reported that the ferromagnetic phase is
orthorhombically distorted at the coexistence region of
$\alpha$-MnAs and $\beta$-MnAs. The existence of such anisotropy in
M111 films could not be deduced from our measurements. This
contribution may be hidden by the demagnetizing and the uniaxial
anisotropy terms. The study of thinner films is needed to define the
role of this contribution in the magnetization of MnAs films.
Corrections of higher order to the in-plane anisotropy of M111
samples are needed to describe the experimental results. These terms
are highly samples dependent, as discussed below.

The uniaxial anisotropy constant, $K_1$ was calculated from the area
enclosed by the loops measured along the easy and the hard
directions. The demagnetization term is calculated taking Ms=870
emu/cm$^3$, as deduced from the magnetization vs. temperature
measurements. The calculated anisotropy constants are $K_1 =
1.3\times10^6$ J/m$^3$ and $K_n= 1.5\times 10^5$ J/m$^3$ for M100
samples and $K_1 = 1.3\times 10^6$ J/m$^3$ for M111 ones. The
uniaxial anisotropy, $K_1$, is of magneto-crystalline origin and
agrees well with the values measured by J. Lindner \emph{et al.}
\cite{Lindner} $K_1$ does not depend notably on the thickness of the
samples or the substrate orientation and is very close to the bulk
value. \cite{Deblois}

\begin{figure}[htb]
\includegraphics[width=\linewidth]{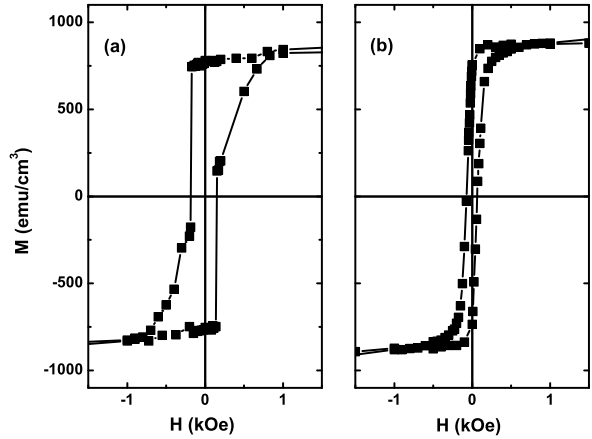}
\caption{Low-field detail of the hysteresis measured with the
applied field oriented along the easy-axis for (a)M100-66, (b)
M111-66. The measurements were performed at 5 K in the SQUID
magnetometer.}\label{Figure3}
\end{figure}

In Fig. \ref{Figure3}, the hysteretic zone of the loops measured
with the magnetic field applied along the in-plane easy-axis is
shown. The shape of the loops of M111 and M100 samples are notably
different. While in M111 samples [Fig. \ref{Figure3}(b)] the
magnetization reversal is smooth, the magnetization in M100 ones
[Fig. \ref{Figure3}(a)] shows a sharp jump at the coercive field
where a partial inversion of the magnetization occurs, while higher
magnetic fields are necessary to completely saturate the
magnetization of the sample. The shapes of the curves are associated
with domains movements and are affected both by defects and other
inhomogeneities and by the anisotropy of the samples. \cite{Jagla}
The coercivity increases as the thickness of the films decreases,
indicating that the magnetization reversal is affected by
substrate-induced strains and topographic defects.

In M100 samples, the coercivity of the loops measured along the
easy-axis, $H_c(0)$ follows the condition:
\begin{equation}
H_c(0) \ll K/M,
\end{equation}
indicating that the magnetization reversal is mainly driven by
domain wall movement.

In M111 samples, the in-plane anisotropy terms are much smaller and
condition (3) is no longer valid. The direction perpendicular to the
film surface is a very strong hard-axis due to the superposition of
the magneto-crystalline anisotropy and the demagnetizing field
resulting in large domain walls.  For this reason we suggest that
the walls in this system are N\'eel-like, i.e., the magnetization
within the walls rotates in the film plane. A simple calculation of
domain wall widths, $\delta$, in M111 films based on the effective
anisotropy of the system results in $\delta> t$ values. According to
the N\'eel criteria, in this case the wall mode using an in-plane
rotation has a lower energy than the classical Bloch
mode.\cite{Hubert}

\begin{figure}[htb]
\includegraphics[width=\linewidth]{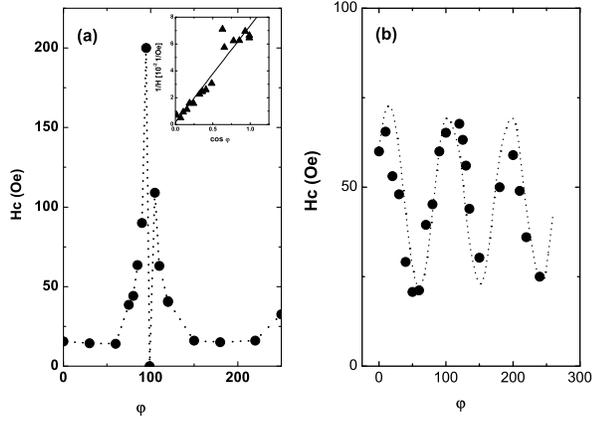}
\caption{Angular dependence of the coercivity, measured with the
field rotating in the plane of the films. (a)M111-66, (b) M100-66.
Measurements performed at 85 K in the VSM.}\label{Figure4}
\end{figure}

The in-plane angular dependence of the magnetization loops has been
carefully measured by VSM. We observe that the magnetization loops
of M100 samples change remarkably as the magnetic field is rotated
in the plane of the films while those of M111 samples vary very
little. The magnetic remanence vs. the azimuthal angle, $\varphi$
curves underlines the symmetry of the magnetic anisotropy of the
films. M100 samples show a strong uniaxial anisotropy with a
hard-axis oriented along [0001] direction. M111 samples, on the
other hand, present a much weaker anisotropy contribution with
different symmetries, depending on the growth conditions. Uniaxial,
three and four-fold anisotropy symmetries have been observed in
different samples. Typical $H_c$($\varphi$) curves for M100 and M111
samples are shown in Fig. \ref{Figure4}.

The coercivity of M100 films shows two high peaks at both sides of
the hard-axis angle. The angular dependence of $H_c$ for $0 <
\varphi < \pi /2$ can be fitted by: \cite{Reich}
\begin{equation}
cos \varphi = H_c(0) / H_c (\varphi)
\end{equation}
suggesting that the magnetization reversal is driven by 180$^o$ wall
movements [inset Fig. \ref{Figure4}(a)]. In a recent article, F.
Schippan and coworkers \cite{SchippanII} report MFM images that show
in similar samples the existence of 180$^o$ walls, oriented along
the $[11\bar{2}0]$ direction. As the uniaxial in-plane anisotropy is
much larger than the demagnetizing term, the walls are Bloch-like,
i.e. the magnetization in the walls rotates out-of-the-plane of the
films. Even if MFM measurements have been performed at room
temperature where the $\alpha$ and the $\beta$ phases coexist, these
results are a strong indication of the domain configuration in these
samples.

The $\varphi$ dependence of the $H_c$ is much weaker for M111
samples and has a four-fold anisotropy in the case of the M111-66
film shown in Fig. \ref{Figure4}(b). The symmetry of the in-plane
anisotropy in these samples can be three-fold and even two-fold and
is strongly dependent on the sample growth parameters: substrate
temperature during sample growth, surface stoichiometry, etc. The
origin of this term may be associated to surface-induced strains
and/or surface reconstruction symmetries. A study of the anisotropy
of M111 ultrathin films is under progress to clarify this point.

\begin{figure}[htb]
\includegraphics[width=\linewidth]{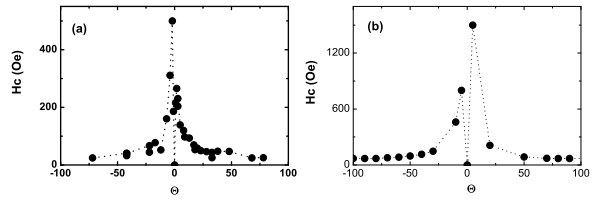}
\caption{Angular dependence of the coercivity measured in the OOP
geometry. (a)M111-66, (b) M100-66. Measurements performed at 85 K in
the VSM.}\label{Figure5}
\end{figure}

The OOP angular dependence of M100 and M111 coercive fields is shown
in Fig. \ref{Figure5}. An important increase of the coercive field
with angle is observed, as the field approaches the film normal
direction. The coercivity abruptly decreases to zero at the
hard-axis of magnetization. M100 and M111 samples display a similar
behavior, following a $H_c(0) / cos (\theta)$ function. The
remanence monotonically decreases to zero as $\theta$ is varied from
$\pi$/2 to 0, keeping $\varphi$ fixed and parallel to  the in-plane
easy axis direction. As condition (3) is fulfilled in this geometry
for both group of samples, the reversal process can be thought as
performed with only the component of the external field parallel to
the easy-axis magnetization direction.

\begin{figure}[htb]
\includegraphics[width=\linewidth]{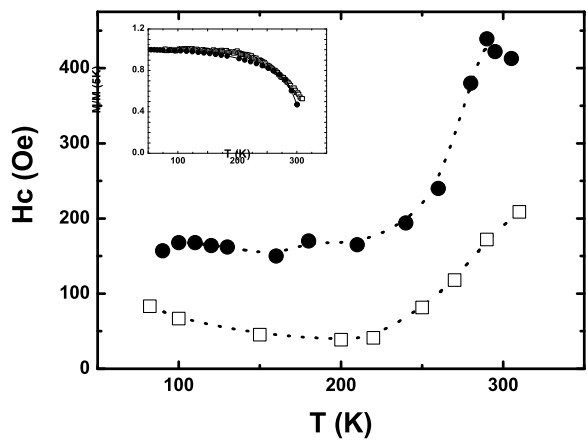}
\caption{Coercive field vs. temperature for ($\Box$) M111-66 and
($\bullet$) M100-66 samples. Data taken from VSM measurements. Inset
of the figure: M vs. T for ($\Box$) M111-100 and the ($\bullet$)
M100-100 samples, respectively. }\label{Figure6}
\end{figure}

The temperature dependence of the coercive field, extracted from
loops measured with the applied field oriented along the in-plane
easy-axis was examined (Fig. \ref{Figure6}). At low temperatures,
T$\leq$200 K, the coercive field decreases smoothly with increasing
temperature, as expected for an ordinary ferromagnet. This behavior
is understood in terms of thermal activation of domain walls.
\cite{Gaunt} At $T_I \sim$200 K, this tendency is reversed and an
increase of the $H_c$ is observed, up to $T_{II} \sim$300 K. This
anomalous behavior is associated to the appearance of new pinning
centers. These new pinning sites would progressively appear as the
temperature increases at nucleation points of the $\beta$-phase or
strained regions of the $\alpha$-phase. Previous papers report the
$\alpha$-$\beta$ phase coexistence in both type of samples but at
higher temperatures. \cite{Daweritz,Mattoso}

The magnetization vs. temperature curves do not show any particular
feature in the T$_I$ - T$_{II}$ range (see Inset of Fig.
\ref{Figure6}). Ferromagnetic resonance measurements indicate that
the magneto-crystalline anisotropy decreases with temperature as the
system approaches the Curie temperature, T$_C$, \cite{nosotros}
without showing noticeable signatures around 200 K. Therefore, the
magneto-crystalline anisotropy of MnAs cannot be the origin of the
observed effect. The M vs. T curve saturates at low temperatures to
the bulk moment \cite{goodenough} but an important increase of the
Curie temperature, T$_C$, from 314 K to 330 K (M100) and 350 K
(M111), is observed. This result suggests that the strains induced
by the GaAs stabilize the $\alpha$-hexagonal phase up to this high
temperature. In fact, X-ray \cite{Mattoso} and neutron
\cite{GarciaIII} diffraction patterns show a $\alpha$-$\beta$
coexistence in the 314 K- T$_C$ temperature range for M111 samples.
Thus, in this temperature range, the magnetization would be given by
the addition of an unsaturated ferromagnetic component that
corresponds to the $\alpha$-phase and a paramagnetic contribution,
arising from the $\beta$ phase. Finally, above room temperature the
thermal energy overcomes the effect of the new pinning centers and
induces a strong reduction of the coercive field as the temperature
approaches T$_C$.

Data shown in Fig. \ref{Figure6} evidence that the reentrant effect
is independent of the substrate orientation. The observed behavior
in the coercivity could then be associated to a former stage of the
$\alpha$-$\beta$ phase coexistence. A. Ney {\it et al.} \cite{Ney}
first observed this anomalous behavior in M100 films and associated
it to the existence of an array of $\alpha$-MnAs stripes. The
authors claim that the temperature dependence of the coercivity
allows them to distinguish between homogeneous and stripes domains
of $\alpha$-MnAs. However, STM images show that M111 samples do not
form ferromagnetic stripes at room temperature. \cite{Ouerghi} Our
finding of a similar behavior in both types of samples suggests that
the origin of this effect does not reside in the appearance of the
$\alpha$-phase in the form of stripes but in the phase coexistence
phenomena, beyond the epitaxy between MnAs and GaAs. The number of
pinning centers is temperature dependent due to the progressive
nucleation of $\beta$-phase regions and/or the strain evolution in
the $\alpha$ phase towards the phase transformation.

\begin{figure}[htb]
\includegraphics[width=\linewidth]{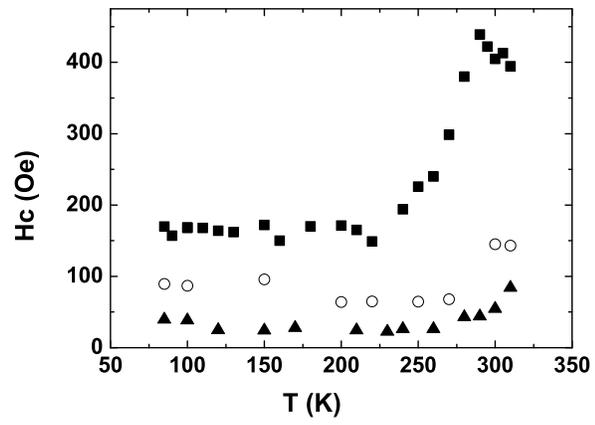}
\caption{Coercive field vs. temperature for ($\blacksquare$)
M100-66, ($\bigcirc$) M100-100 and ($\blacktriangle$) M100-200
samples. Data taken from VSM measurements.}\label{Figure7}
\end{figure}

Fig. \ref{Figure7} shows the temperature dependence of the
coercivity for M100 samples of different thicknesses.  The increase
of the $H_c$ with temperature, in the T$_I$ - T$_{II}$ zone is more
pronounced for thinner films. However, the functional form of the
temperature variation of the coercivity does not depend on
thickness.

The results indicate, in summary, that even though the coercivity in
this case is essentially given by extrinsic properties, the
reentrant behavior seems to be an intrinsic characteristic of the
magnetism of MnAs thin films.

\section{CONCLUSIONS}
We have shown that the magnetization reversal in MnAs films depends
on the substrate orientation and is strongly affected by the
magneto-crystalline anisotropy of the compound.

On one hand, the angular dependence of the coercive field of films
grown on GaAs(100) shows that the in-plane magnetization reversal
occurs by 180$^0$ domain wall movements. Moreover, the shape of the
hysteresis loops indicates that the magnetization alignment with the
magnetic field is driven in two-steps. This is explained by the
existence of a distribution of potential barriers for domain
movement, that have a very sharp threshold at the coercive field
where more than 70\% of the magnetization turns into the direction
of the applied field. The magnetization alignment is strongly
affected by the in-plane hard uniaxial anisotropy of the system,
that induces the formation of Bloch walls along the films.

On the other hand the uniaxial anisotropy of samples grown on
GaAs(111)B is oriented perpendicular to the film surface, leading to
the creation of N\'{e}el domain walls and a smooth reversal of the
magnetization in a low-field range.

Summarizing, we have demonstrated that the coercivity of the films
presents a general anomalous behavior as a function of temperature,
independent of the substrate orientation and films thicknesses. This
effect is associated to the appearance of new pinning centers as a
consequence of the progressive growth of a $\beta$-phase or
distortion of the $\alpha$-phase as the temperature is increased
above 200 K.

\section{ACKNOWLEDGEMENTS}
 The authors thank partial financial support from Fundaci\'on Antorchas, FONCyT
 03-13297, Universidad Nacional de Cuyo, CONICET and the ECOS-SUD.

\clearpage


\end{document}